\documentclass[hyper]{JHEP} 
\usepackage{epsfig}
\usepackage{amsmath}
 \usepackage{slashed}

\newcommand\fverb{\setbox\pippobox=\hbox\bgroup\verb}
\newcommand\fverbdo{\egroup\medskip\noindent%
            \fbox{\unhbox\pippobox}\ }

\newcommand\fverbit{\egroup\item[\fbox{\unhbox\pippobox}]}
\newbox\pippobox
\title{$SL(2, Z)$ invariant rotating $(m,n)$ strings in $AdS_3\times S^3$ with mixed flux}

\preprint{arXiv:1610.03402 [hep-th]}

\author{Sorna Prava Barik$^a$,
Malak Khouchen$^b$, Josef Kluso\v{n}$^b$ and Kamal L. Panigrahi$^a$\\
$^a$ Department of Physics, Indian Institute of Technology
Kharagpur,
Kharagpur-721 302, India\\

$^b$ Institute for Theoretical Physics  and Astrophysics, Faculty
of Science,\\ Masaryk University, Kotl\'{a}\v{r}sk\'{a} 2, 611 37,
Brno,  Czech Republic\\ E-mail: \email{sornaprava15@gmail.com,
malak.khouchen@gmail.com, klu@physics.muni.cz,
panigrahi@phy.iitkgp.ernet.in}}

\abstract{We study rigidly rotating and pulsating $(m,n)$ strings in
$AdS_3 \times S^3$ with mixed three form   flux. The $AdS_3 \times
S^3$ background with mixed three form  flux is obtained in the near
horizon limit of $SL(2,Z)$-transformed solution, corresponding to
the bound state of NS5-branes and fundamental strings. We study the
probe $(m,n)-$string in this background by solving the manifest
$SL(2,Z)-$covariant form of the action. We find out the dyonic giant
magnon and single spike solutions corresponding to the equations of
motion of a probe string in this background and find out various
relationships among the conserved charges. We further study a class
of pulsating $(m,n)$ string in $AdS_3$ with mixed three form  flux.}
\keywords{AdS/CFT correspondence, Semiclassical strings}
\def\det{\mathrm{det}}

\def\bm{\mathbf{m}}
\def\mM{\mathcal{M}}

\def\bB{\mathbf{B}}

\newcommand{\hg}{\hat{g}}

\newcommand{\mJ}{\mathcal{J}}

\begin{document}
\section{Introduction}
Integrability in string theory has been proved to be one of the
most useful techniques in studying string spectrum in various
semisymmetric superspaces \cite{Zarembo:2010sg} \footnote{for a
detailed introduction and references on integrability in AdS/CFT
refer \cite{Beisert:2010jr}.}. The appearance of integrability on
both sides of the AdS/CFT correspondence \cite{Maldacena:1997re,
Witten:1998qj, Gubser:1998bc} has added tremendous amount of
progress in the study of string theory. In this context, type IIB
superstring theory on $AdS_5\times S^5$ has been shown to be
described as supercoset sigma model \cite{Metsaev:1998it}. The
appearance of integrability via appearance of hidden charges was
first exploited in \cite{Bena:2003wd}. With the realization that
the counting of gauge invariant operators from gauge theory side
can be elegantly formulated in terms of an integrable spin chain,
it has been established that integrability played an important
role on both sides of the duality, since the dual string theory is
integrable in the semiclassical limit. In this connection, a
special limit was put forth using in which both sides of the
duality were analyzed in great detail. In particular, the spectrum
on the field theory side was shown to consist of  elementary
excitations, the so called magnons which carry momentum $p$ along
the finitely or infinitely long spin chain. On the string theory
side, the dual string state derived from the rigidly rotating
string in the ${\cal R}\times S^3$ appears to give the same
dispersion relation between the string energy (E) and the angular
momentum (J) in the large 't Hooft limit and is known as the giant
magnon\cite{Hofman:2006xt}. A more general kind of rotating
strings, known as spiky strings, are dual to higher twist
operators also presented in \cite{Kruczenski:2004wg}. It was
further argued that they both fall into the category of special
class of general rotating string solutions \cite{Frolov:2003qc} on
the sphere \cite{Ishizeki:2007we}. In addition to the rigidly
rotating strings, the spinning and pulsating strings have also
been shown to have exact correspondence with some dual operators
in the gauge theory\cite{Minahan:2002rc}. Pulsating string was
introduced first in \cite{Gubser:2002tv}. Compared to the rigidly
rotating strings, the folded and the pulsating strings are less
studied even though the pulsating-rotating solutions offer better
stability than the non pulsating solutions \cite{Khan:2005fc}.
These solutions are time-dependent as opposed to the usual rigidly
rotating string solutions. They are expected to be dual to highly
excited states in terms of operators \cite{Gubser:2002tv}.

Long ago, Schwarz \cite{Schwarz:1995dk} has constructed an
$SL(2,Z)$ multiplet of string-like solutions in type IIB string
theory starting from the fundamental string solution. It is known
that the equations of motion of type IIB supergravity theory are
invariant under an $SL(2, R)$ group. This suggests the possibility
to generate new supergravity solutions by applying this rotation
to known solutions such as string-like as well as five-brane
solutions. A discrete subgroup $SL(2,Z)$ of this $SL(2, R)$ group
has been conjectured later to be the exact symmetry group of the
type IIB string theory based on the fact that there are no
fractional string or D-brane charges. The $SL(2,Z)$ transformed
solution of the a bound state of $Q5$ NS5-branes and $Q1$
fundamental strings (F-strings) is characterized by  charges with
respect to RR and NS-NS two forms. In the near horizon limit of
this solution, we obtain the $AdS_3 \times S^3$ background with
mixed three form fluxes with integer charges. It has also been
recently shown that the $SL(2,Z)$-transformation and the near
horizon limit commute. This allows to map the $(m,n)-$string in
$AdS_3 \times S^3$ background with mixed three form fluxes to
$(m',n')-$string in $AdS_3 \times S^3$ background with NS-NS two
form flux. Recently in a series of papers
\cite{Hoare:2013pma}\cite{Hoare:2013ida}\cite{Hoare:2013lja} the
superstring theory on $AdS_3 \times S^3 \times T_4$ supported by a
combination of RR and NSNS 3-form fluxes (with parameter of the
NSNS 3-form $q$ ) has been investigated in detail. The worldsheet
theory interpolates between the pure RR flux model $(q = 0)$ and
the pure NSNS flux model $(q = 1)$ \footnote{For a nice review and
comprehensive list of references on the study of integrability of
superstrings on $AdS_3\times S^3 \times T^4$ with both RR and
mixed flux refer \cite{Sfondrini:2014via}}. The theory has been
shown to be integrable and for a generic value of the parameter
$q$ the corresponding tree-level S-matrix for massive BMN-type
excitations has been computed. Further computations along the
lines of rotating and pulsating string have been studied, for
example
in\cite{David:2014qta}\cite{Banerjee:2014gga}\cite{Borsato:2014hja}\cite{Hernandez:2014eta}
\cite{Lloyd:2014bsa}\cite{Stepanchuk:2014kza}
\cite{Banerjee:2015bia}\cite{Kluson:2015lia}\cite{Banerjee:2015qeq}\cite{Banerjee:2016avv}.
In view of the study of superstrings in $AdS_3\times S^3$
background with mixed flux, it is interesting to investigate
further the rotating $(m, n)$ string in $AdS_3\times S^3$
background with mixed three form fluxes. This problem can be
mapped to $(m', n')$ string in $AdS_3\times S^3$ background with
NS-NS two form flux by using the symmetries of the intersecting
brane background itself. The rest of the paper is organized as
follows. In section 2, we study $(m, n)- $string in $AdS_3 \times
S^3$ background with mixed three form fluxes after mapping it to
the simpler $ (m',n')- $string in $AdS_3 \times S^3$ background
with NS-NS two form flux. In section 3, we solve the corresponding
equations of motion in the single angular momentum case where we
find out solutions that correspond to spike and giant magnon.
Section 4 is devoted to the study of the rotating string with two
angular momenta and we present the relations among various
conserved charges. In section 5, we discuss the pulsating strings
in AdS$_3$ background with mixed three form flux. In section 6, we
conclude and present our outlook.

\section{Rotating $(m,n)-$string in $AdS_3\times S^3$ with
mixed flux}\label{second} We begin this section with a review of
the construction of $AdS_3\times S^3$ background with mixed three
form fluxes that was performed recently in \cite{Kluson:2016dca}.
The starting point is the  $AdS_3\times S^3$ background with NS-NS
two form \footnote{We ignore the part of the metric corresponding
to four torus $T^4$ with the volume $V_4$.}
\begin{eqnarray}\label{AdS3S3NSNS}
ds^2
&=&L^2[ds^2_{\widetilde{AdS}_3}+ds^2_{\Omega^3}] \ , \nonumber \\
H&=&2L^2 (\tilde{\epsilon}_{AdS_3}+\epsilon_{S^3}) \ ,  L^2=r^2_5 \ ,
\nonumber \\
e^{-2\Phi_{NS}}&=&\frac{1}{g_s^2}\frac{r_1^2}{r_5^2} \ , r_1^2=\frac{16
\pi^4 g_s^2 \alpha'^3 Q_1}{V_4} \ , r^2_5=Q_5
 \alpha' \ , V_4=(2\pi)^4\alpha'^2 v \ ,  \nonumber \\
\end{eqnarray}
 where $ds^2_{\widetilde{AdS}_3}$
is the line element of $AdS_3$ space expressed in dimensionless
variables. It is well known that the solution given in
(\ref{AdS3S3NSNS}) is a  solution of type IIB supergravity
equations of motion. On the other hand, we also know that type IIB
superstring theory possesses $SL(2,Z)-$duality transformation that
leaves the metric in the Einstein frame unchanged. In case of two
forms, it is convenient to introduce the vector $\bB$ defined as
 \begin{equation}\label{bBSL}
\bB=\left(\begin{array}{cc} B \\
C^{(2)} \\ \end{array}\right)
\end{equation}
where $B$ and $C^{(2)}$ are NSNS and RR two forms respectively. The
vector $\bB$ transforms under $SL(2,Z)$ transformation as
\begin{equation}
\hat{\bB}=(\Lambda^T)^{-1} \bB \ ,
\end{equation}
where
\begin{equation}\label{Lambdadef}
\Lambda=\left(\begin{array}{cc} a & b \\
c & d \\ \end{array}\right) \ , \det \Lambda=1
 \ ,
 \end{equation}
 and where $a,b,c,d$ are integers. Type IIB theory also has
 two scalar fields $\chi$ and $\Phi$, where the dilaton $\Phi$
 is in the NS-NS sector while $\chi$ belongs to the RR sector.
 It is convenient to combine these fields  into a
 complex field $\tau=\chi+ie^{-\Phi}$ and introduce the following matrix
 \begin{equation}
\mM=e^\Phi\left(\begin{array}{cc} \tau \tau^* & \chi \\
\chi & 1 \\
\end{array}\right)=e^\Phi\left(\begin{array}{cc}\chi^2+e^{-2\Phi} & \chi
\\
\chi & 1 \\ \end{array}\right) \ , \det \mM=1
\end{equation}
that transforms under $SL(2,Z)$ transformation as
\begin{equation}\label{mMSL}
\hat{\mM}=\Lambda \mM \Lambda^T \ ,
\end{equation}
where $\Lambda$ is given in (\ref{Lambdadef}).

Then in order to find $AdS_3\times S^3$ background with mixed
three form fluxes, we perform $SL(2,Z)$ transformation of the
ansatz (\ref{AdS3S3NSNS}) and we obtain the line element  in the
form  \cite{Kluson:2016dca}
\begin{equation}
d\bar{s}^2=\sqrt{\frac{c^2}{g_s^2}\frac{r_1^2}{r_5^2}+d^2}
\left[L^2[ds^2_{\widetilde{AdS}_3}+ds^2_{\Omega^3}]+ds_T^2\right] \ ,
\end{equation}
where $ds^2_T=dx_6^2+\dots +dx_9^2$. We see that the new solution
has the curvature radius
$\bar{L}^2=\sqrt{\frac{c^2}{g_s^2}\frac{r_1^2}{r_5^2}+d^2}L^2=
\frac{1}{g_s}\sqrt{c^2 r_1^2+d^2g_s^2 r^2_5}r_5$. Further, there are the
following NSNS and RR three forms
\begin{eqnarray}
\tilde{H}&=&dH=2dL^2 (\tilde{\epsilon}_{AdS_3}+\epsilon_{S^3}) \ ,
\nonumber \\
\tilde{F}&=&-bH= - 2bL^2 (\tilde{\epsilon}_{AdS_3}+\epsilon_{S^3})
\nonumber \\
\end{eqnarray}
and dilaton and zero form RR field as
\begin{eqnarray}
e^{-\tilde{\Phi}}&=&
\frac{\sqrt{Q_1 Q_5 v}}{c^2 Q_1+d^2 Q_5 v}=\frac{1}{\hg_s}  \
,
\nonumber \\
\tilde{\chi}&=&
\frac{acQ_1+bdvQ_5}{c^2 Q_1+d^2 Q_5v} \ . \nonumber \\
\end{eqnarray}
Our goal is to study the dynamics of the probe $(m,n)-$string in
this background.

To do this, we introduce the action for $(m,n)$-string that has the
form \footnote{For recent discussion, see \cite{Kluson:2016pxg}.}
\begin{eqnarray}\label{mnstringaction}
S_{(m,n)}&=&-T_{D1}\int d\tau d\sigma\sqrt{\bm^T\mM^{-1} \bm
}\sqrt{-\det g_{MN}\partial_\alpha x^M\partial_\beta x^N} +
\nonumber \\
&+&T_{D1}\int d\tau d\sigma \bm^T \bB_{MN}\partial_\tau
x^M\partial_\sigma x^N \ ,\nonumber \\
\end{eqnarray}
where
\begin{equation}
\bm=\left(\begin{array}{cc} m \\
n \\ \end{array}\right) \ , \quad  \bm^T \mM^{-1}\bm=\bm_i
(\mM^{-1})^{ij}\bm_j \ , \bm^T\bB=\bm_i \bB^i \ ,
\end{equation}
where  $m,n$ count the number of fundamental string $(m)$ and
D1-branes $(n)$ and hence they have to be integers.

It is important that the action (\ref{mnstringaction}) is manifestly
invariant under $SL(2,Z)$ transformation when $\bB$ and $\mM$
transform as in (\ref{bBSL}) and (\ref{mMSL}) and when $\bm$
transforms as
\begin{equation}
\hat{\bm}=\Lambda \bm \ .
\end{equation}
Note that this action is  expressed using Einstein frame metric
$g_{MN}$ which is related to the string frame metric $G_{MN}$ by
the relation $g_{MN}=e^{-\Phi/2}G_{MN}$ where the Einstein frame
metric is invariant under $SL(2,Z)$-transformations. Then it was
shown in \cite{Kluson:2016dca} that $(m,n)-$string in mixed
$AdS_3\times S^3$ background with mixed three form fluxes can be
mapped to $(m',n')-$string in $AdS_3\times S^3$ background with
NSNS three form flux. Explicitly, using the manifest covariance of
the action (\ref{mnstringaction}) we obtain
\begin{eqnarray}\label{mnprobeact}
S_{(m,n)}
&=&-T_{D1}\int d\tau d\sigma \sqrt{m'^2+n'^2e^{-2\Phi_{NS}}}
\sqrt{-\det (G_{MN}\partial_\alpha x^M\partial_\beta x^N)}+
\nonumber \\
&+&T_{D1}\int d\tau d\sigma m'B_{MN}\partial_\tau x^M\partial_\sigma
x^N \ ,
\nonumber \\
\end{eqnarray}
where  $G_{MN},B_{MN}$ and $\Phi^{NS}$ correspond to the background
(\ref{AdS3S3NSNS})
and we used the fact that
\begin{equation}
\bm^T \mM^{-1}\bm=(\Lambda^{-1} \bm)^T \mM_{NS}^{-1}
(\Lambda^{-1}\bm)=
(m'^2+n'^2
e^{-2\Phi_{NS}}) e^{\Phi_{NS}}
\end{equation}
with
\begin{equation}\label{bmprime}
\bm'=\left(\begin{array}{cc} m' \\
n'\\ \end{array}\right)=\left(\begin{array}{cc} dm -bn \\
-cm+an \\ \end{array}\right) \ .
\end{equation}
We see that we reduced the problem of the dynamics of $(m,n)-$string
in mixed $AdS_3\times S^3$ background to the much simpler analysis
of $(m',n')-$string in pure NSNS $AdS_3\times S^3$ background where
the action is given in  (\ref{mnprobeact}). On the other hand, this
action is non-linear due to the presence of the square root of the
determinant that makes the analysis of equations of motion rather
awkward.
For that reason it is useful to rewrite this action into Polyakov-like form when we introduce
an auxiliary metric $\gamma_{\alpha\beta}$ and write the action
$S_{(m,n)}$ into the form
\begin{equation}\label{SmnPolyakov}
S=-\frac{\tau_{(m,n)}}{2}\int d\tau d\sigma
\sqrt{-\gamma}\gamma^{\alpha\beta}
G_{\alpha\beta}+q_{(m,n)}\int d\tau d\sigma B_{MN}
\partial_\tau x^M \partial_\sigma x^N \ ,
\end{equation}
where
\begin{equation}
\tau_{(m,n)}=T_{D1}\sqrt{m'^2+n'^2e^{-2\Phi_{NS}}} \ , \quad
q_{(m,n)}=T_{D1}m'=T_{D1}\bm^T \mathbf{q}\ ,
\end{equation}
where $\mathbf{q}=\left(\begin{array}{cc} d \\
-b \\ \end{array}\right)$ is the charge vector of $(d,-b)-$flux
background. Note that we have also used the fact that $\Phi_{NS}$
is constant for the background (\ref{AdS3S3NSNS}). To see an
equivalence between (\ref{SmnPolyakov}) and Nambu-Goto form of the
action, note that the equations of motion for
$\gamma_{\alpha\beta}$ have the form
\begin{eqnarray}\label{gammaeq}
T_{\alpha\beta}=-\frac{2}{\sqrt{-\gamma}}
\frac{\delta S_{(m,n)}}{\delta \gamma^{\alpha\beta}}=\tau_{(m,n)}[-
\gamma_{\alpha\beta}\gamma^{\gamma\delta}G_{\gamma\delta}+2G_{\alpha\beta}]=0
\nonumber \\
\end{eqnarray}
that has clearly a solution
$\gamma_{\alpha\beta}=G_{\alpha\beta}$. Inserting this solution
into (\ref{SmnPolyakov}) we obtain the original action. In the
following we use the Polyakov form of the action due to the
manifest linearity of the theory. The equations of motion with
respect to $\gamma$ have been already determined in
(\ref{gammaeq}) while the equations of motion with respect to
$x^M$ can be easily determined from (\ref{SmnPolyakov})
\begin{eqnarray}\label{eqPolyakov}
&-&\frac{\tau_{(m,n)}}{2}\sqrt{-\gamma}\gamma^{\alpha\beta}\partial_M
G_{KL}\partial_\alpha x^K \partial_\beta x^L+\tau_{(m,n)}\partial_\alpha[\sqrt{-\gamma}\gamma^{\alpha\beta}G_{MN}
\partial_\beta x^N]+\nonumber \\
&+&q_{(m,n)}H_{MNK}\partial_\tau x^N\partial_\sigma x^K=0 \ , \nonumber \\
\end{eqnarray}
where
\begin{equation}
H_{MNK}=\partial_M B_{NK}+\partial_N B_{KM}+\partial_K B_{MN} \ .
\end{equation}
Our goal is to find solutions of the equations of motion derived
above that correspond to giant magnon or single spike
configurations. For that reason it is convenient to use the
following explicit form of the background metric
(\ref{AdS3S3NSNS})
\begin{eqnarray}
ds^2&=&L^2[-\cosh^2\rho dt^2+
d\rho^2+\sinh^2\rho d\phi^2+d\theta^2+\sin^2\theta d\phi_1^2+\cos^2\theta d\phi^2_2] , \nonumber \\
& &b_{t\phi}=L^2\sinh^2\rho, \quad b_{\phi_1\phi_2}=-L^2\cos^2\theta , \nonumber \\
\end{eqnarray}
where we used ordinary symbols for coordinates instead of symbols
with tilde used in (\ref{AdS3S3NSNS}) keeping in mind that all
coordinates are dimensionless. Note that due to the fact that the
action does not depend explicitly on $\phi,\phi_1,\phi_2$ and $t$,
the action is invariant under constant shifts
\begin{equation}
t'(\sigma^\alpha)=t(\sigma^\alpha)+\epsilon_t \ , \quad \phi'(\sigma^\alpha)=
\phi(\sigma^\alpha)+\epsilon_\phi \ , \quad
\phi'_{1,2}(\sigma^\alpha)=\phi_{1,2}(\sigma^\alpha)+\epsilon_{1,2} \ ,
\end{equation}
where all $\epsilon's$ are constants. With the help of the standard
Noether theorem we derive the following conserved currents
\begin{eqnarray}
\mJ^\alpha_t&=&-\tau_{(m,n)}\sqrt{-\gamma}G_{tM}\partial_\beta x^M
\gamma^{\beta\alpha}+q_{(m,n)}B_{tN}\epsilon^{\alpha\beta}\partial_\beta x^N \ ,
\nonumber \\
\mJ^\alpha_\phi&=&-\tau_{(m,n)}\sqrt{-\gamma}G_{\phi M}\partial_\beta x^M
\gamma^{\beta\alpha}+q_{(m,n)}B_{\phi N}\epsilon^{\alpha\beta}\partial_\beta x^N \nonumber \\
\mJ^\alpha_{\phi_1,\phi_2}&=&-\tau_{(m,n)}\sqrt{-\gamma}G_{\phi_{1,2} M}\partial_\beta x^M
\gamma^{\beta\alpha}+q_{(m,n)}B_{\phi_{1,2} N}\epsilon^{\alpha\beta}\partial_\beta x^N \nonumber \\
\end{eqnarray}
where $\epsilon^{\tau\sigma}=-\epsilon^{\sigma\tau}=1$. Note that these
currents obey the relations
\begin{equation}
\partial_\alpha \mJ^\alpha_A=0 \ , A=t,\phi,\phi_{1,2} \ .
\end{equation}
Using these relations we derive the
following conserved charges
\begin{eqnarray}\label{charges}
P_t&=&\int d\sigma \left(
-\tau_{(m,n)}\sqrt{-\gamma}G_{tt}\partial_\beta t
\gamma^{\beta\tau}+q_{(m,n)}B_{t\phi}\epsilon^{\tau\sigma}\partial_\sigma  \phi \right) \ ,\nonumber \\
J_\phi&=&
\int d\sigma \left(
-\tau_{(m,n)}\sqrt{-\gamma}G_{\phi \phi}\partial_\beta \phi
\gamma^{\beta\tau}+q_{(m,n)}B_{\phi t}\epsilon^{\tau\sigma}\partial_\sigma t
\right) \ , \nonumber \\
J_{\phi_1}&=&
\int d\sigma \left(
-\tau_{(m,n)}\sqrt{-\gamma}G_{\phi_1 \phi_1}\partial_\beta \phi_1
\gamma^{\beta\tau}+q_{(m,n)}B_{\phi_1\phi_2 }\epsilon^{\tau\sigma}\partial_\sigma \phi_2
\right) \ , \nonumber \\
J_{\phi_2}&=&
\int d\sigma \left(
-\tau_{(m,n)}\sqrt{-\gamma}G_{\phi_2 \phi_2}\partial_\beta \phi_2
\gamma^{\beta\tau}+q_{(m,n)}B_{\phi_2\phi_1 }\epsilon^{\tau\sigma}\partial_\sigma \phi_1
\right) \ . \nonumber \\
\end{eqnarray}
Now we try to solve the equations of motion explicitly when we
consider the following ansatz
\begin{eqnarray}
t=\gamma\tau, \quad \theta=\theta(y), \quad
\phi_1=\omega_1\tau+g_1(y), \quad   \phi_2=\omega_2\tau+g_2(y) \ ,
\end{eqnarray}
where $y$ is a function of world sheet coordinates  $y = \alpha\sigma + \beta\tau$, together with $\rho=0$ and $\phi=0$.
At the same time we impose the conformal gauge when $\gamma_{\tau\tau}=-1 \ ,
\gamma_{\sigma\sigma}=1 \ , \gamma_{\tau\sigma}=0$. In this case the components
of the stress energy tensor have the form
\begin{eqnarray}\label{gammaeqh}
T_{\tau\tau}&=&T_{\sigma\sigma}=\tau_{(m,n)}[G_{\tau\tau}+G_{\sigma\sigma}]=0
\ ,
\nonumber \\
T_{\tau\sigma}&=&2\tau_{(m,n)}G_{\tau\sigma}=0 \ , \nonumber \\
\end{eqnarray}
where
\begin{eqnarray}
& &G_{\tau\tau}=\partial_\tau x^M \partial_\tau x^N
 G_{MN}=L^2[-\gamma^2+
 \beta^2\theta'^2+(\omega_1+\beta {g'}_1)^2\sin^2\theta+
 (\omega_2+\beta {g'}_2)^2\cos^2\theta] \ , \nonumber \\
& &G_{\sigma\sigma}=
\partial_\sigma x^M \partial_\sigma x^N G_{MN}=
\alpha^2L^2[\theta'^2+{g'}_1^2\sin^2\theta+{g'}_2^2\cos^2\theta] \ , \nonumber \\
& &G_{\tau\sigma}=g_{\sigma\tau}=L^2[\alpha\beta\theta'^2
+\alpha(\omega_1+\beta {g'}_1){g'}_1\sin^2\theta+
\alpha(\omega_2+\beta {g'}_2){g'}_2\cos^2\theta ] \ . \nonumber \\
\end{eqnarray}
The equation of motion for $\phi_1$ implies
\begin{eqnarray}\label{gprime1}
g'_1
=\frac{1}{(\alpha^2-\beta^2)}
\left(\frac{\Phi_1}{\sin^2\theta}+\beta \omega_1-
\frac{q_{(m,n)}}{\tau_{(m,n)}}\alpha\omega_2 \right) \ , \nonumber \\
\end{eqnarray}
where $\Phi_1=\mathrm{const}$. and $g_1'=
\frac{\partial g_1}{\partial y}$ 
\\
In the same way the equation of motion for $\phi_2$ implies
\begin{eqnarray}
g'_2
=\frac{1}{(\alpha^2-\beta^2)}
\left(\frac{\Phi_2}{\cos^2\theta}+\beta\omega_2-\frac{q_{(m,n)}}{\tau_{(m,n)}}\alpha
\omega_1 \right) \ . \nonumber \\
\end{eqnarray}
From these two equations we can see one important point that the case of the single
angular momentum, i.e., $\omega_2=0 \ , g_2=0$ is possible only  when $q_{(m,n)}=0$
as follows from the equation of motion for $\phi_2$.

In order to find the equation of motion for $\theta$ we use the constraint
$T_{\tau\tau}=0$ and we obtain
\begin{eqnarray}\label{Ttautau}
&-&L^2\gamma^2(\alpha^2-\beta^2)^2+L^2\sin^2\theta\omega_1^2(\alpha^2-\beta^2)^2+
L^2\cos^2\theta\omega_2^2(\alpha^2-\beta^2)^2+\nonumber \\
&+&L^2(\alpha^2+\beta^2)(\alpha^2-\beta^2)^2
\theta'^2+2L^2\beta\omega_1(\alpha^2-\beta^2)\sin^2\theta
(\frac{\Phi_1}{\sin^2\theta}+\beta\omega_1+\frac{q_{(m,n)}}{\tau_{(m,n)}}\alpha\omega_2)+
 \nonumber \\
 &+&2L^2\beta\omega_2(\alpha^2-\beta^2)\cos^2\theta
 (\frac{\Phi_2}{\cos^2\theta}+\beta\omega_2-\frac{q_{(m,n)}}{\tau_{(m,n)}}\alpha\omega_1)+
 \nonumber \\
 &+&L^2\sin^2\theta(\alpha^2+\beta^2)
 (\frac{\Phi_1}{\sin^2\theta}+\beta\omega_1+\frac{q_{(m,n)}}{\tau_{(m,n)}}\alpha\omega_2)^2
 +\nonumber \\
&+&L^2\cos^2\theta(\alpha^2+\beta^2)
 (\frac{\Phi_2}{\cos^2\theta}+\beta\omega_2-\frac{q_{(m,n)}}{\tau_{(m,n)}}\alpha\omega_1)^2
 =0 \ . \nonumber \\
\end{eqnarray}
This equation simplifies considerably when we impose the boundary
condition   that for $\theta\rightarrow \frac{\pi}{2} ,\theta'=0$. Since $\lim_{\theta\rightarrow \frac{\pi}{2}}\cos\theta=0$,  we have to demand that $\Phi_2=0$ and
the previous equation implies
\begin{eqnarray}\label{eqgammagen}
&-&\gamma^2(\alpha^2-\beta^2)^2+\omega_1^2(\alpha^2-\beta^2)^2+
2\beta\omega_1(\alpha^2-\beta^2)(
\Phi_1+\beta\omega_1-\frac{q_{(m,n)}}{\tau_{(m,n)}}\alpha\omega_2)+
 \nonumber \\
 &+&(\alpha^2+\beta^2)
 (\Phi_1+\beta\omega_1-\frac{q_{(m,n)}}{\tau_{(m,n)}}\alpha\omega_2)^2
 =0 \nonumber \\
\end{eqnarray}
that can be solved for $\gamma$.
Let us now consider the constraint $T_{\tau\sigma}=0$ that implies
\begin{eqnarray}\label{Ttausigmazero}
G_{\tau\sigma}=
g_{\theta\theta}\alpha\beta \theta'^2+g_{\phi_1\phi_1}\alpha g'_1(\omega_1+
\beta g'_1)+g_{\phi_2\phi_2}\alpha g'_2(\omega_2+\beta g'_2)=0
 \nonumber \\
\end{eqnarray}
that for $\theta=\pi/2, \theta'(\pi/2)=0$ implies
\begin{equation}
g'_1|_{\theta=\pi/2}(\omega_1+\beta g'_1|_{\theta=\pi/2})=0
\end{equation}
and we have to analyze under which condition this equation is
obeyed. The first possibility is that $g'_1|_{\theta=\pi/2}=0$ and
using (\ref{gprime1}) we find that this is possible when
\begin{equation}\label{PhiI1}
\Phi^{I}_1=-\beta\omega_1\textbf{+}\frac{q_{(m,n)}}{\tau_{(m,n)}}\alpha\omega_2
\ .
\end{equation}
The second possibility how to obey (\ref{Ttausigmazero}) is to
demand that  $\omega_1+\beta g'_1|_{\theta=\pi/2}=0$ which implies
\begin{equation}\label{PhiII1}
\Phi^{II}_1=-\frac{1}{\beta}(\alpha^2\omega_1
-\beta\frac{q_{(m,n)}}{\tau_{(m,n)}}\alpha\omega_2)
\end{equation}
These two values of the constants $\Phi^{I}_1,\Phi^{II}_2$ determine whether
we have giant spike or giant magnon solution. Before we proceed to the
discussion of the general case with two angular momenta, we consider the simpler
case of single angular momentum.
%
%
%
%
%
%
%
\section{Single Angular Momentum}
Let us now consider the case when
$g'_2=0$ and $\omega_2=0$. As we argued previously this is possible on
condition when $q_{(m,n)}=0$ too. In this case we have
\begin{equation}
g'_1=\frac{1}{(\alpha^2-\beta^2)}
\left(\frac{\Phi_1}{\sin^2\theta}+\beta \omega_1 \right) \
\end{equation}
while the constraint $T_{\tau\tau}=0$ takes the form
\begin{eqnarray}
&-&L^2\gamma^2(\alpha^2-\beta^2)^2+L^2\sin^2\theta\omega_1^2(\alpha^2-\beta^2)^2+\nonumber \\
&+&L^2(\alpha^2+\beta^2)(\alpha^2-\beta^2)^2
\theta'^2+2L^2\beta\omega_1(\alpha^2-\beta^2)\sin^2\theta
(\frac{\Phi_1}{\sin^2\theta}+\beta\omega_1)+
 \nonumber \\
  &+&L^2\sin^2\theta(\alpha^2+\beta^2)
 (\frac{\Phi_1}{\sin^2\theta}+\beta\omega_1)^2
 =0 \ . \nonumber \\
\end{eqnarray}
If we impose again the condition   that for $\theta=\frac{\pi}{2},\theta'=0$
we obtain the equation
\begin{equation}\label{eqsingle}
-\gamma^2(\alpha^2-\beta^2)^2+\omega_1^2(\alpha^2-\beta^2)^2+
2\beta\omega_1(\alpha^2-\beta^2)(\Phi_1+\beta \omega_1)+
(\alpha^2+\beta^2)(\Phi_1+\beta\omega_1)^2=0
\end{equation}
that can be solved for $\gamma$. Further, the constraint
$T_{\tau\sigma}=0$ has the form
\begin{eqnarray}
G_{\tau\sigma}=
g_{\theta\theta}\alpha\beta \theta'^2+g_{\phi_1\phi_1}\alpha g'_1(\omega_1+
\beta g'_1)=0
 \nonumber \\
\end{eqnarray}
that for $\theta=\pi/2, \theta'(\pi/2)=0$ implies
\begin{equation}
g'_1|_{\theta=\pi/2}(\omega_1+\beta g'_1|_{\theta=\pi/2})=0
\end{equation}
that can be solved for two values of $\Phi_1^{I,II}$
\begin{equation}
\Phi^{I}_1=-\beta\omega_1\ ,
\end{equation}
and
\begin{equation}
\Phi^{II}_1=-\frac{\alpha^2}{\beta}\omega_1 \ .
\end{equation}
We begin with the first case.
\subsection{Giant Magnon Solution}
We first consider the case with $\Phi^{I}_1=-\beta\omega_1$ . Equation (\ref{eqsingle}) implies $\gamma=\omega_1$
while  the constraint $T_{\tau\tau}=0$ gives the equation for
$\theta'$
\begin{eqnarray}
\theta'^2=\dfrac{\omega_1^2
\cos^2\theta(\alpha^2\sin^2\theta-\beta^2)}{(\alpha^2-\beta^2)^2\sin^2\theta} \ . \nonumber \\
\end{eqnarray}
Using (\ref{charges}), we find the explicit form of the conserved charges
\begin{eqnarray}
P_t&=&
-2\kappa L^2 \tau_{(m,n)}\int {d\theta
\dfrac{(\alpha^2-\beta^2)\sin\theta}{\alpha\cos\theta\sqrt{\alpha^2\sin^2\theta-\beta^2}}}
 \ , \nonumber \\
J_{\phi_1}&=&
2\kappa L^2\tau_{(m,n)} \int {d\theta \dfrac{\sin\theta}{\alpha
\cos\theta}\sqrt{\alpha^2\sin^2\theta-\beta^2}} \ ,
\end{eqnarray}
where $\kappa$ counts the number of spikes on the $(m,n)-$string
world-volume.
 Both of these integrals diverge but
the difference between $E=-P_t$ and $J_{\phi_1}$  is finite and is
equal to
\begin{eqnarray}\label{giantmagnondisprela}
E-J_{\phi_1}=2\kappa\tau_{(m,n)}
L^2\sqrt{1-\left(\frac{\beta}{\alpha}\right)^2}=2\kappa\tau_{(m,n)}
L^2\sin(\dfrac{\triangle \phi_1}{2}) \ ,
\end{eqnarray}
where we introduced the difference  angle $\phi_1$ defined as
\begin{eqnarray}
\triangle\phi_1 &=& 2 \int_{\phi^{min}_1}^{\phi^{max}_1}d\phi_1
=2\alpha\int_{\theta_{min}}^{\theta^{max}}{\dfrac{{g'}_1d\theta}{\alpha\mid\theta'\mid}}
=-2 \arccos\left(\frac{\beta}{\alpha}\right) \ .
\end{eqnarray}
Note that (\ref{giantmagnondisprela}) is the generalization of the
giant magnon dispersion relation to the specific case of
$(m,n)-$string in $(d,-b)-$mixed flux background. More explicitly,
the condition $q_{(m,n)}=T_{D1}m'=0$ implies that $dm=bn$ and
$\tau_{(m,n)}=T_{D1}e^{-\Phi_{NS}}n'=T_{D1}\frac{n}{d}e^{-\Phi_{NS}}$.
The special case is when $n'=1$ that corresponds to $(b,d)-$string
and we derive giant magnon dispersion relation. Note that this is in
agreement with the $SL(2,Z)-$covariance of Type IIB string theory
since $(d,-b)$-flux background is derived from $(1,0)-$flux
background by $SL(2,Z)-$rotation with the matrix
$\Lambda=\left(\begin{array}{cc} a & b \\
c & d \\ \end{array}\right)$ where NSNS and RR two forms transform
as
\begin{equation}
\left(\begin{array}{cc} B'\\
C'^{(2)} \\ \end{array}\right)=(\Lambda^T)^{-1}\
\left(\begin{array}{cc} B\\
C^{(2)} \\ \end{array}\right)
\end{equation}
while $(m,n)$-string transforms as $\bm'=\Lambda \bm$, so that
$(b,d)-$string is $SL(2,Z)$ rotation of D1-brane.
\subsection{Spike Solution}
In the second case, we have $\Phi^{II}_1=-\frac{\alpha^2}{\beta}\omega_1$. Equation (\ref{eqsingle}) gives
$\gamma^2=\frac{\alpha^2}{\beta^2}\omega_1^2$ while  the constraint
$T_{\tau\tau}=0$ implies following differential equation for
$\theta$
\begin{eqnarray}
\theta'^2=\dfrac{\alpha^2\omega_1^2\cos^2\theta
(\beta^2\sin^2\theta-\alpha^2)}{\beta^2(\alpha^2-\beta^2)^2\sin^2\theta} \ . \nonumber \\
\end{eqnarray}
It is easy to see that the energy is equal to
\begin{eqnarray}
E&=&\dfrac{2\kappa L^2 \tau_{(m,n)}}{\alpha}\int {d\theta \dfrac{(\alpha^2-\beta^2)\sin\theta}{\cos\theta\sqrt{\beta^2\sin^2\theta-\alpha^2}}}
%
\end{eqnarray}
which is divergent. On the other hand note that $J_{\phi_1}$ is now
finite and is equal to
\begin{eqnarray}
 J_{\phi_1}&=& 
 -2\kappa\tau_{(m,n)} L^2\cos\left(\frac{\alpha}{\beta}\right) \ . \nonumber \\
\end{eqnarray}
In order to find finite dispersion relation, let us determine the
angle difference
\begin{eqnarray}
\triangle\phi_1 
&=&\frac{2}{\alpha} \int_{\theta_{min}}^{\theta^{max}} {d\theta \dfrac{\sqrt{(\beta^2\sin^2\theta-\alpha^2)}}{\sin\theta\cos\theta}}
\nonumber \\
\end{eqnarray}
that is divergent.
Then it is easy to find following dispersion relation
\begin{eqnarray}
E+\kappa\tau_{(m,n)}L^2\triangle \phi_1
=2\kappa\tau_{(m,n)}L^2\left(\frac{\pi}{2}-\theta_1\right) \ .  \nonumber \\
\end{eqnarray}
This is the dispersion relation corresponding to the single spike
solution of the string.
\section{Two Angular Momenta}
In this section we consider a more interesting case of two non-zero
angular momenta. Recall that in section (\ref{second}) we determined
that these solutions are characterized by condition $\Phi_2=0$ and
two values of $\Phi_1$ given in (\ref{PhiI1}) and (\ref{PhiII1}).
Let us begin with the first case.
\subsection{First Limiting Case $\Phi^{I}_1$}
We begin with the first case with $\Phi^{I}_1=-\beta\omega_1+\frac{q_{(m,n)}}{\tau_{(m,n)}}\alpha\omega_2$. Note that for this value of $\Phi^I_1$, the equation
(\ref{eqgammagen}) implies $\gamma=\omega_1$.  \\

Using (\ref{Ttautau}), we get the differential equation of $\theta$
\begin{eqnarray}
\theta'^2
&=&\dfrac{\Omega^2\cos^2\theta}{(\alpha^2-\beta^2)^2\sin^2\theta}
[\sin^2\theta-\sin^2\theta_0 ] \ , \nonumber \\
\end{eqnarray}
where $\Omega^2=\alpha^2(1-\frac{q^2_{(m,n)}}
{\tau^2_{(m,n)}})(\omega_1^2-\omega_2^2)$ and
$\sin\theta_0=\dfrac{(\beta\omega_1-\alpha\omega_2\frac{q_{(m,n)}}{\tau_{(m,n)}})}{\Omega}$.\\

The explicit form of the conserved charge
$E$ is
\begin{eqnarray}
E
&=&\dfrac{2\kappa\omega_1\tau_{(m,n)}L^2(\alpha^2-\beta^2)}{\alpha\Omega}\int{\dfrac{d\theta
\sin\theta}{\cos\theta \sqrt{\sin^2\theta-\sin^2\theta_0 }}}
\end{eqnarray}
which is divergent. As for the remaining conserved charges $J_{\phi_1}$ and
$J_{\phi_2}$, we get
\begin{eqnarray}
J_{\phi_1}
&=&\dfrac{2\kappa\omega_1\tau_{(m,n)}
L^2}{\alpha\Omega}(\alpha^2-\beta^2) \int{ d\theta
\dfrac{\sin\theta}
{\cos\theta\sqrt{\sin^2\theta-\sin^2\theta_0}}}+\nonumber \\
&-&\dfrac{2\kappa\alpha\omega_1\tau_{(m,n)} L^2}{\Omega}
\left(1-\frac{q^2_{(m,n)}}{\tau^2_{(m,n)}}\right) \int{ d\theta
\dfrac{\sin\theta\cos\theta}{\sqrt{\sin^2\theta-\sin^2\theta_0}}}
\nonumber \\
\end{eqnarray}
and
\begin{eqnarray}
J_{\phi_2}
&=&\dfrac{2\kappa\tau_{(m,n)} L^2\alpha\omega_2}{\Omega} \left(
1-\frac{q^2_{(m,n)}}{\tau^2_{(m,n)}}\right)
\int d\theta{ \dfrac{{\sin\theta\cos\theta}}{\sqrt{\sin^2\theta-\sin^2\theta_0}}} +\nonumber \\
&-&{2\kappa\tau_{(m,n)}
L^2\frac{q_{(m,n)}}{\tau_{(m,n)}}\sin\theta_0}\int
d\theta{\dfrac{\cos\theta}{\sin\theta\sqrt{\sin^2\theta-\sin^2\theta_0}}}
\end{eqnarray}
and the angle difference
\begin{eqnarray}
\triangle \phi_1
&=&\dfrac{2(\alpha\omega_2\dfrac{q_{(m,n)}}{\tau_{(m,n)}}-\beta\omega_1)}{\Omega}\int{\dfrac{d\theta \cos\theta}{\sin\theta\sqrt{\sin^2\theta-\sin^2\theta_0}}}\nonumber \\
&=&-2\sin\theta_0\int{\dfrac{d\theta \cos\theta}{\sin\theta\sqrt{\sin^2\theta-\sin^2\theta_0}}}\nonumber \\
&=&-2\cos^{-1}(\sin\theta_0) \ . \nonumber \\
\end{eqnarray}
Collecting these results together we obtain the following dispersion
relation
\begin{eqnarray}
\dfrac{E}{\omega_1}-\dfrac{J_{\phi_1}}{\omega_1}
=\dfrac{J_{\phi_2}}{\omega_2}-\dfrac{\kappa
q_{(m,n)}L^2\triangle\phi_1}{\omega_2} \ . \nonumber \\
\end{eqnarray}
Using previous integral we evaluate the right side of the equation
above and we obtain
\begin{eqnarray}
\frac{J_{\phi_2}-\kappa q_{(m,n)}L^2\triangle \phi_1}{\omega_2}=
\frac{1}{\omega_1}\sqrt{(J_{\phi_2}-\kappa q_{(m,n)}L^2\triangle \phi_1)^2+
4\kappa^2\tau^2_{(m,n)}L^4
\left(1-\frac{q^2_{(m,n)}}{\tau^2_{(m,n)}}\right)(1-\sin^2\theta_0)}
\nonumber \\
\end{eqnarray}
and hence we derive the final form of the dispersion relation
\begin{eqnarray}
& &E-J_{\phi_1} =\sqrt{(J_{\phi_2}-\kappa q_{(m,n)}L^2\triangle \phi_1)^2+
4\kappa^2\tau^2_{(m,n)}L^4
\left(1-\frac{q^2_{(m,n)}}{\tau^2_{(m,n)}}\right)\sin^2\frac{\triangle
\phi_1}{2}}
\nonumber \\
&=&\sqrt{(J_{\phi_2}-\kappa \bm^T \mathbf{q}T_{D1}L^2\triangle \phi_1)^2+
4\kappa^2\bm^T \mM^{-1}\bm e^{-\Phi_{NS}}T_{D1}^2L^4
\left(1-\frac{\bm^T \mathbf{q}}{\bm^T \mM^{-1}\bm
e^{-\Phi_{NS}}}\right)\sin^2\frac{\triangle
\phi_1}{2}} \ . \nonumber \\
\end{eqnarray}
This dispersion relation is the generalization of the dispersion
relations derived in \cite{Hoare:2013lja} and also in
\cite{Banerjee:2014gga} to the case of $(m,n)-$string in
$(d,-b)-$mixed flux background. We see that this dispersion relation
is linear in the $\triangle \phi_1$ that is identified with the
world-sheet momentum $p$ which spoils periodicity of this solution.
On the other hand it is clear that this dispersion relation reduces
to the usual giant magnon dispersion relation when $\bm^T
\mathbf{q}=0$ and also $n'=1$ that corresponds to $(b,d)-$string in
$(d,-b)-$flux background and the dispersion relation has the form
\begin{equation}
E-J_{\phi_1}=\sqrt{J^2_{\phi_2}+4\kappa^2
e^{-2\Phi_{NS}}L^4T^2_{D1}\sin^2\frac{\triangle\phi_1}{2}} \
\end{equation}
that again corresponds to $SL(2,Z)-$rotation of the dispersion
relation of D1-brane in pure NSNS flux background. On the other hand
it is interesting to analyze dispersion relation when $n'=0$ that
implies $q_{(m,n)}=\tau_{(m,n)}$ that corresponds to
$m=\frac{an}{c}$. If we again consider the case when $m'=1$ we find
that this corresponds to $(a,c)-$string and we find that the
dispersion relation has the form
\begin{equation}
E-J_{\phi_1}=J_{\phi_2}-\kappa T_{D1}L^2 \triangle \phi_1  \ .
\end{equation}
We interpret this solution as the bound state of $J_{\phi_2}$
elementary magnons so that for $J_{\phi_2}=1$ we obtain massless
dispersion relation
\begin{equation}
E-J_{\phi_1}=\epsilon=1-\kappa T_{D1}L^2\triangle_{\phi_1}
\end{equation}
that has nice physical interpretation. The $(a,c)-$string in
$(d,-b)-$flux background is defined by $SL(2,Z)-$rotation of
fundamental string in pure NSNS flux background,  where the matrix
$\Lambda$ is given in (\ref{Lambdadef}). On the other hand we know
that fundamental string in $AdS_3\times S^3$ with NSNS flux has
exact WZW conformal field theory description with massless
dispersion relation \cite{Hoare:2013lja}.

\subsection{Second Limiting Case $\Phi^{II}_1$}
Now, we consider the second case with $\Phi^{II}_1=-\frac{1}{\beta}(\alpha^2\omega_1
 -\beta\frac{q_{(m,n)}}{\tau_{(m,n)}}\alpha\omega_2)$. From (\ref{eqgammagen}) we again find
$\gamma^2=\dfrac{\alpha^2}{\beta^2}\omega_1^2$\\

Using (\ref{Ttautau}) we find that the
 differential equation for
 $\theta$ has the form
\begin{eqnarray}\label{thetaPhiII}
\theta'^2
&=&\dfrac{\Omega^2\cos^2\theta}{(\alpha^2-\beta^2)^2\sin^2\theta}[\sin^2\theta-\sin^2\theta_0
] \ ,
\end{eqnarray}
where $
\Omega^2=\alpha^2\left(1-\frac{q^2_{(m,n)}}{\tau^2_{(m,n)}}\right)(\omega_1^2-\omega_2^2)
\ , \quad
\sin\theta_0=\frac{1}{\beta\Omega}\left(\alpha^2\omega_1-\alpha\beta\omega_2\dfrac{q_{(m,n)}}{\tau_{(m,n)}}\right)
 \ .$

Using (\ref{thetaPhiII}) we determine the following conserved charges
\begin{eqnarray}\label{EPhiII}
E
&=&\dfrac{2\kappa\omega_1\tau_{(m,n)}L^2(\alpha^2-\beta^2)}{\beta\Omega}\int{\dfrac{d\theta\sin\theta}
{\cos\theta\sqrt{\sin^2\theta-\sin^2\theta_0}} } \ , \nonumber \\
J_1
&=&{-2\kappa\omega_1\tau_{(m,n)} L^2}
\sqrt{\dfrac{1-\frac{q^2_{(m,n)}}{\tau^2_{(m,n)}}}
{\omega_1^2-\omega_2^2}}\cos\theta_0
 \ , \nonumber \\
J_2
&=&{2\kappa
\tau_{(m,n)} L^2\omega_2}\sqrt{\dfrac{1-\frac{q^2_{(m,n)}}
{\tau^2_{(m,n)}}}
{\omega_1^2-\omega_2^2}}
\cos\theta_0 -2n\tau L^2\frac{q_{(m,n)}}{\tau_{(m,n)}}
(\frac{\pi}{2}-\theta_0) \ \nonumber \\
\end{eqnarray}
together with the angle difference
\begin{eqnarray}\label{trianglephiII}
& &\triangle \phi_1
=-2\sin\theta_0\int_{\theta_{min}}^{\pi/2}
\frac{d\theta\cos\theta}{\sin\theta\sqrt{\sin^2\theta-\sin^2\theta_0}}-2\dfrac{\omega_1}{\beta\Omega}(\alpha^2-\beta^2)\int{\dfrac{d\theta\sin\theta}{\cos\theta\sqrt{\sin^2\theta-\sin^2\theta_0}}} \nonumber \\
\end{eqnarray}
From (\ref{EPhiII}) and (\ref{trianglephiII}) we see that $E$
and $\triangle\phi_1$ are both divergent  but their combination is
finite and is equal to
\begin{eqnarray}
E+\kappa\tau_{(m,n)} L^2\triangle\phi_1&=&
2\kappa\tau_{(m,n)}
L^2 (\frac{\pi}{2 }-\theta_0) \ . \nonumber \\
\end{eqnarray}
Further, the dispersion relation between angular momenta can be
written as
\begin{eqnarray}\label{J1J2}
J_1=\sqrt{(2\kappa\tau_{(m,n)}
L^2)^2(1-\frac{q^2_{(m,n)}}{\tau^2_{(m,n)}})\sin^2\dfrac{(\triangle\phi_1)_{reg}}{2}+[J_2-\kappa\tau_{(m,n)}
L^2\frac{q_{(m,n)}}{\tau_{(m,n)}}(\triangle\phi_1)_{reg}]^2}\nonumber\\
\end{eqnarray}
where $(\triangle\phi_1)_{reg}=-2\cos^{-1}(\sin\theta_0)$. Note that
(\ref{J1J2}) could be rewritten using the original variables $\bm$
and $\mM$ and we could discuss the various properties of this
relation with dependence on the values of vector $\bm$ exactly in
the same way as in previous section but we will not repeat it here
since the discussion would be exactly the same.

\section{ Pulsating $(m,n)-$string in $AdS_3$ with mixed flux}
In this section we will analyze the pulsating $(m,n)-$string in
mixed three form flux background. Recall that such a string has an
action
\begin{equation}\label{SmnPolyakovpuls}
S=-\frac{\tau_{(m,n)}}{2}\int d\tau d\sigma
\sqrt{-\gamma}\gamma^{\alpha\beta}
G_{\alpha\beta}+q_{(m,n)}\int d\tau d\sigma B_{MN}
\partial_\tau x^M \partial_\sigma x^N \ ,
\end{equation}
with the equation of motion for  $x^M$
\begin{eqnarray}\label{eqxmpuls}
&-&\frac{\tau_{(m,n)}}{2}\sqrt{-\gamma}\gamma^{\alpha\beta}\partial_M
G_{KL}\partial_\alpha x^K \partial_\beta x^L+\tau_{(m,n)}\partial_\alpha[\sqrt{-\gamma}\gamma^{\alpha\beta}G_{MN}
\partial_\beta x^N]+\nonumber \\
&+&q_{(m,n)}H_{MNK}\partial_\tau x^N\partial_\sigma x^K=0 \ , \nonumber \\
\end{eqnarray}
and  the equations of motion for $\gamma_{\alpha\beta}$
\begin{eqnarray}\label{gammaeqpuls}
T_{\alpha\beta}=-\frac{2}{\sqrt{-\gamma}}
\frac{\delta S_{(m,n)}}{\delta \gamma^{\alpha\beta}}=\tau_{(m,n)}[-
\gamma_{\alpha\beta}\gamma^{\gamma\delta}G_{\gamma\delta}+2G_{\alpha\beta}]=0 \ .
\nonumber \\
\end{eqnarray}
In order to find pulsating $(m,n)-$string in $AdS_3$, we consider the
following ansatz
\begin{eqnarray}
t = t(\tau),\quad  \rho = \rho (\tau),  \quad  \phi = \kappa\sigma
\end{eqnarray}
with $b_{t\phi}=L^2\sinh^2\rho$. Then it is easy to find the form of the
induced metric
\begin{eqnarray}
G_{\tau\tau}&=&
L^2(-\cosh^2\rho \dot{t}^2+\dot{\rho}^2) \ , \quad
G_{\sigma\sigma}=
L^2 \kappa^2 \sinh^2\rho  \ , \quad G_{\tau\sigma}=G_{\sigma\tau}=0 \ .
\nonumber \\
\end{eqnarray}
so that (\ref{gammaeqpuls}) in the conformal gauge takes the form
\begin{eqnarray}\label{gammaeqosc}
& &T_{\tau\tau}=T_{\sigma\sigma}=
\tau_{(m,n)}L^2[-\cosh^2\rho \dot{t}^2+\dot{\rho}^2+ \kappa^2 \sinh^2\rho]=0 \ ,
\nonumber \\
& &T_{\tau\sigma}=2\tau_{(m,n)}G_{\tau\sigma}=0 \ . \nonumber \\
\end{eqnarray}
From (\ref{eqxmpuls}) we find that the equation of motion for $t$
has very simple form
\begin{eqnarray}\label{eqtime}
& &L^2(\tau_{(m,n)}
\dot{t}\cosh^2\rho
-\kappa q_{(m,n)}\sinh^2\rho) =\mathcal{E} \ . \nonumber \\
\end{eqnarray}
On the other hand the equation of motion for $\rho$ is more
complicated and is equal to
\begin{eqnarray}
&-&\frac{\tau_{(m,n)}}{2}(-\partial_\rho
G_{tt}\dot{t}^2+\kappa^2\partial_\rho G_{\phi\phi})-L^2\tau_{(m,n)}
\ddot{ \rho}
+2\kappa L^2 q_{(m,n)}\sinh\rho\cosh\rho\dot{t}=0 \ , \nonumber \\
\end{eqnarray}
while the equation of motion for $\phi$ implies
\begin{eqnarray}
\kappa L^2\tau_{(m,n)}\sinh^2\rho
=A \ . \nonumber \\
\end{eqnarray}
Now using (\ref{eqtime}) we find that $P_t$ is equal to
\begin{eqnarray}
P_t=-E
&=&-\int d\sigma L^2 \left(
\tau_{(m,n)}\cosh^2\rho \dot{t}-\kappa q_{(m,n)}\sinh^2\rho\right)
=-2\pi\mathcal{E} \ . \nonumber \\
\end{eqnarray}
The second important quantity is the oscillation
number that is  associated with string motion along the $\rho$ direction
\begin{eqnarray}
N=\oint{d\rho \Pi_{\rho}} \ ,
\end{eqnarray}
where $\Pi_\rho$ is the canonical momentum conjugate to $\rho$
\begin{eqnarray}
\Pi_\rho&=&\tau_{(m,n)}L^2\dot{\rho} \ . \nonumber \\
\end{eqnarray}
Now using the equation (\ref{gammaeqosc}) and (\ref{eqtime})
we obtain differential equation for $\rho$ in the form
\begin{eqnarray}
\dot{\rho}^2
&=&\dfrac{(\kappa q_{(m,n)}L^2\sinh^2\rho+\mathcal{E})^2-\kappa^2\tau^2_{(m,n)} L^4\sinh^2\rho\cosh^2\rho}{\tau^2_{(m,n)} L^4\cosh^2\rho} \nonumber \\
\end{eqnarray}
and hence the oscillation number $N$ is equal to
\begin{eqnarray}
N
&=& \oint{\dfrac{d\rho}{{\cosh\rho}} {\sqrt{(\kappa q_{(m,n)}
L^2\sinh^2\rho+\mathcal{E})^2-\kappa^2\tau^2_{(m,n)} L^4\sinh^2\rho\cosh^2\rho}}} \nonumber
\end{eqnarray}
Changing the variable $x=\sinh\rho$, we get
\begin{eqnarray}
N&=&\oint{\dfrac{dx}{{1+x^2}} {\sqrt{(\kappa q_{(m,n)}
L^2x^2+\mathcal{E})^2-\kappa^2\tau^2_{(m,n)} L^4x^2(1+x^2)}}} \nonumber\\
&=& \kappa L^2 \sqrt{\tau_{(m,n)}^2-q_{(m,n)}^2}\int_0^{\sqrt{R_+}}{\dfrac{dx}{1+x^2}\sqrt{(R_+ -x^2)(x^2-R_-)}}\nonumber \\
&=& \kappa L^2 \sqrt{\tau_{(m,n)}^2-q_{(m,n)}^2}\dfrac{1}{\sqrt{-R_-}}\left[R_- E\left(\dfrac{R_+}{R_-}\right)+(1+R_+)\left[K\left(\dfrac{R_+}{R_-}\right)-(1+R_-)\Pi\left(-R_+,\dfrac{R_+}{R_-}\right)\right]\right] \ , \nonumber \\
\end{eqnarray}

where $R_+$ and $R_-$ are the roots of the quadratic equation in the square root with

\begin{eqnarray}
R_{\pm}=\dfrac{2q_{(m,n)}\mathcal{E}-\kappa L^2 \tau^2_{(m,n)} \pm \tau_{(m,n)}\sqrt{-4L^2\kappa q_{(m,n)}\mathcal{E}+4\mathcal{E}^2 +L^4 \kappa^2 \tau^2_{(m,n)}}}{2\kappa L^2 (\tau^2_{(m,n)}-q^2_{(m,n)})}.
\end{eqnarray}
In short string limit
\begin{eqnarray}
N&=&\frac{\pi }{4 L^2 \kappa \tau_{(m,n)}} \mathcal{E}^2 +\frac{\pi
q_{(m,n)} }{4 L^4 \kappa^2 \tau^3_{(m,n)}}
\mathcal{E}^3-\frac{5\pi(-3q^2_{(m,n)} +\tau^2_{(m,n)}) } {32 L^6
\kappa^3 \tau^5_{(m,n)}} \mathcal{E}^4\nonumber \\
&-&\frac{7 \pi q_{(m,n)}
 (-5q^2_{(m,n)}+3\tau^2_{(m,n)}) }
 {32 L^8 \kappa^4 \tau^7_{(m,n)}}\mathcal{E}^5+O[\mathcal{E}]^6 \ .
\end{eqnarray}
Reversing the above series, we have the expression for $\mathcal{E}$
\begin{eqnarray}
\mathcal{E}&=&\frac{2 \sqrt{L^2 \kappa \tau_{(m,n)}}}{\sqrt{\pi}
}\sqrt{N}-\frac{2 q_{(m,n)} }{\pi \tau_{(m,n)}} N+\frac{5
(-q^2_{(m,n)}+\tau^2_{(m,n)})}{2 \pi^{3/2} \tau^2_{(m,n)} \sqrt{L^2
\kappa \tau_{(m,n)}} } N^{3/2} \nonumber \\
&+&\frac{6 (-q^3_{(m,n)}+q_{(m,n)} \tau^2_{(m,n)})}{L^2 \kappa \pi^2
\tau^4_{(m,n)}} N^2 + O[N]^{5/2} \ .
\end{eqnarray}
Putting $m'=1$ and $n'=0$, we have the following expression 
\begin{eqnarray}
N=\frac{\pi }{4 L^2 \kappa T_{D1}}\mathcal{E}^2 +\frac{\pi  }{4 L^4 \kappa^2  T_{D1}^2}\mathcal{E}^3+\frac{5\pi }{16 L^6 \kappa^3  T_{D1}^3}\mathcal{E}^4+\frac{7 \pi }{16 L^8 \kappa^4  T_{D1}^4}\mathcal{E}^5+O[\mathcal{E}]^6 \nonumber
\end{eqnarray}
\begin{eqnarray}
\mathcal{E}=\frac{2 \sqrt{L^2 \kappa T_{D1}}}{\sqrt{\pi} }\sqrt{N}-\frac{2  }{\pi } N + O[N]^{5/2}
\end{eqnarray}
The condition $n'=0$ implies that $cm=an$ and
$\tau_{(m,n)}=T_{D1}m'=T_{D1}\frac{n}{c}$.
Again putting $m'=1$, we get $m=a$ and $n=c$. So we note that the $(a,c)-$string in $(d,-b)$-flux background is the $SL(2,Z)$ rotation of $F-$string and  our result for the specific case $m'=1$ and $n'=0$ should match with \cite{Banerjee:2015bia}.
\section{Conclusion}
In this paper, we have studied the rotating and pulsating
$(m,n)-$type string in $AdS_3 \times S^3$ background with mixed
fluxes which has been obtained by taking the $SL (2,Z)$
transformations of the usual $(F1-NS5)$ bound state followed by a
near horizon geometry. We have applied $SL(2,Z)$ transformation on
the $(m,n)-$probe string action and generated $(m',n')-$string
action, where the $m'$ and $n'$ are the $SL (2, Z)$ invariants.
The giant magnon and its dyonic counterpart solutions have been
obtained by solving relevant equations of motion of the probe
string in the above background in the presence of NS-NS fluxes. We
have shown various regularized dispersion relations among different
conserved charges that the background admits. We have also checked
that in the absence of probe D-string charges, the relations among
various charges do match exactly with the F-string result. Furthermore, we have
 looked at an oscillating $(m, n)-$string in the background
of $AdS_3$ with NS-NS flux. In short string limit, we have obtained
the energy of such a string in terms of the oscillation number. The
work done in this paper can be extended in several ways. One of
the interesting problems to consider is to study the pulsating and
circular string solutions of $(m, n)-$type string in the $R\times
S^3$ with the NS-NS flux turned on. A point to note, however, is
that the pulsating and oscillating strings in $R\times S^3$ is
qualitatively different from the $AdS_3$ case. It is left as a
further example for future work. In the context of obtaining the
mixed flux background, one of the backgrounds which one might look
for is the $AdS_3 \times S^3 \times S^3 \times S^1$ with mixed
flux. A way to do so would be to start with the
$NS1-NS1'-NS5-NS5'$ intersecting brane solution of
\cite{Papadopoulos:1999tw} and then apply the $SL (2,Z)$
transformation followed by a rotation and check the commutativity
of the these two operations. At present, it appears to be a nice
idea to pursue further.

\vskip .5in \noindent {\bf Acknowledgement:}
\\
 The work of J.K. and M.K. was
supported by the Grant Agency of the Czech Republic under the grant
P201/12/G028.

\end{document}